\documentclass[conference]{IEEEtran}

\usepackage{cite}
\usepackage{amsmath,amssymb}
\usepackage{interval}
\usepackage{hyperref}
\usepackage{cool}
\usepackage{breqn}
\usepackage{wasysym}
\usepackage{dsfont}
\usepackage{graphicx}
\usepackage{times, epsfig}
\usepackage{subfig}
\usepackage{verbatim}
\usepackage[latin1]{inputenc}
\newtheorem{thm}{Theorem}

\newtheorem{defn}{Definition}

\newtheorem{rk}{Remark}
\newtheorem{cor}{Corollary}

\usepackage[usenames, dvipsnames]{color}

\newlength{\figwidth}
\begin{document}
	\title{Detection of a Signal in Colored Noise: A Random Matrix Theory Based Analysis}
	
\author{
\IEEEauthorblockN{
Lahiru~D.~Chamain\IEEEauthorrefmark{1},
Prathapasinghe~Dharmawansa\IEEEauthorrefmark{2}, 
Saman~Atapattu\IEEEauthorrefmark{3}, and 
Chintha~Tellambura\IEEEauthorrefmark{4}
}
\IEEEauthorblockA{
\IEEEauthorrefmark{1}Department of Electrical and Computer  Engineering, University of California Davis, CA 95616 \\
\IEEEauthorrefmark{2}Department of Electronic and Telecommunication Engineering, University of Moratuwa, Moratuwa, Sri Lanka \\
\IEEEauthorrefmark{3}Department of Electrical and Electronic Engineering, University of Melbourne, Victoria, Australia \\
\IEEEauthorrefmark{4}Department of Electrical and Computer Engineering, University of Alberta, Edmonton, Canada
}
}

	\maketitle
	
	

	%
    
    \begin{abstract}
    This paper investigates the classical statistical signal processing problem of detecting a signal in the presence of colored noise with an unknown covariance matrix. In particular, we consider a scenario where 
    $m$-dimensional $p$ possible signal-plus-noise samples and $m$-dimensional $n$ noise-only samples are available at the detector. Then the presence of a signal can be detected using the largest generalized eigenvalue (l.g.e.) of the so called whitened sample covariance matrix. This amounts to statistically characterizing the maximum eigenvalue of the deformed Jacobi unitary ensemble (JUE). To this end, we employ the powerful orthogonal polynomial approach to determine a new finite dimensional expression for the cumulative distribution function (c.d.f.) of the l.g.e. of the deformed JUE. This new c.d.f. expression facilitates the further analysis of the receiver operating characteristics (ROC) of the detector. It turns out that, for $m=n$, when $m$ and $p$ increase such that $m/p$ is fixed, there exists an optimal ROC profile corresponding to each fixed signal-to-noise ratio (SNR). In this respect, we have established a tight approximation for the corresponding optimal ROC profile.
    \end{abstract}

	\IEEEpeerreviewmaketitle

\section{Introduction}

The detection of an unknown noisy signal or a transmit node  is the fundamental task in  many signal processing  and wireless communication applications \cite{Nadakuditi2008sp,Nadakuditi2010jsacsp,Debbah2011it,Couillet2013it,Nadakuditi2017it}. For instance, 
the state-of-the-art of cognitive radio or radar and sonar systems identify the presence of the primary user activity or the existence of the target based on certain statistical properties of the observation vector \cite{Debbah2011it}. 
Among all detection techniques, the sample eigenvalue (of the sample covariance matrix) based detection has gained   prominence recently (see \cite{Bianchi2011tit} and references therein). In this context, the largest sample eigenvalue, also known as the Roy's largest root test, has been popular among detection theorists. Under the common  Gaussian setting with white noise, this amounts to determine the largest eigenvalue of a Wishart matrix having a so-called spiked covariance (see \cite{Baik2005annprob,Baik2006jma} and references therein). 

Certain practical scenarios give rise to additive correlated noise (also known as colored noise) \cite{Maris2003biomed,Vinogradova2013sp,Hiltunen2015sp,Nadakuditi2017it}. Based on the assumption that one has access to signal-plus-noise sample covariance matrix and noise only sample covariance matrix, Rao and Silverstein \cite{Nadakuditi2010jsacsp} proposed a framework to use the generalized eigenvalues of the whitened 
signal-plus-noise sample covariance matrix for detection. The assumption of having the noise only sample covariance matrix is realistic in many practical situations as detailed in \cite{Nadakuditi2010jsacsp}. The  fundamental {\it high dimensional} limits of the generalized sample eigenvalue based detection in colored noise have  been thoroughly investigated in  \cite{Nadakuditi2010jsacsp}. However, to our best knowledge, a tractable {\it finite dimensional} analysis is not available in the literature. Thus,  in this paper, we characterize the statistics of the Roy's largest root in the finite dimensional colored noise setting. 
The Roy's largest root of the generalized eigenvalue detection problem in the Gaussian setting amounts to finite dimensional characterization of the largest eigenvalue of the deformed Jacobi ensemble. Various asymptotic expressions for the Roy's largest root have been derived in \cite{Johnstone2017Biometrika,Dharmawansa2014arx,Dharmawansa2014arx2,wang2017stat} for deformed Jacobi ensemble. However, finite dimensional expressions are available for Jacobi ensemble only (i.e., without the deformation) \cite{koev2005distribution,Dumitriu2012phys}. Although finite dimensional, these expressions are not amenable to further manipulations. Therefore, in this paper, we present simple and tractable closed-form solution to the cumulative distribution function (c.d.f.) of the maximum eigenvalue of the deformed Jacobi ensemble. This expression further facilitates the analysis of the receiver operating characteristics (ROC) of the Roy's largest root test. All these results are made possible due to a novel alternative joint eigenvalue density function that we have derived based on the contour integral approach due to \cite{Dharmawansa2014stat,ONATSKI2014rmt,Passemier2015jmulti,Mo2012,meWang}. 

The key results developed in this paper enable us to understand the joint effect of the system dimensionality ($m$), the number of samples available from the signal-plus-noise ($p$) and noise-only ($n$)  observations, and the signal-to-noise ratio ($\gamma$) on the ROC. For instance, the relative disparity between $m$ and $n$ improves the ROC profile for fixed values of the other parameters. However, the general finite dimensional ROC expressions turns out to give little analytical insights. Therefore, in view of obtaining more insights, we have particularly focused on the case for which the system dimensionality equals the number of samples available from the noise-only observations (i.e., $m=n$). Since this equality is the minimum requirement for the validity of the  whitening operation, from the ROC perspective, it corresponds to the worst possible case when then other parameters being fixed. It turns out that, under the above scenario, when $m$ and $p$ increase such that $m/p$ is fixed, there exists an optimal ROC profile. 
Therefore, the above insight can be of paramount importance in designing future wireless communication systems (i.e., 5G and beyond) with massive degrees of freedom. 

The following notation is used throughout this paper. The superscript $(\cdot)^\dagger$ indicates the Hermitian transpose, $\text{det}(\cdot)$ denotes the determinant of a square matrix, $\text{tr}(\cdot)$ represents the trace of a square matrix, and $\text{etr}(\cdot)$ stands for $\exp\left(\text{tr}(\cdot)\right)$. The $n\times n$ identity matrix is represented by $\mathbf{I}_n$ and the Euclidean norm of a vector $\mathbf{w}$ is denoted by $||\mathbf{w}||$. A diagonal matrix with the diagonal entries $a_1,a_2,\ldots, a_n$ is denoted by $\text{diag}(a_1, a_2,\ldots,a_n)$. We denote the $m\times m$ unitary group by $U(m)$. Finally, we use the following notation to compactly represent the
determinant of an $n\times n$ block matrix:
\begin{equation*}
\begin{split}
\det\left[a_{i}\;\; b_{i,j}\right]_{\substack{i=1,2,\ldots,n\\
 j=2,3,\ldots,n}}&=\left|\begin{array}{ccccc}
 a_{1} & b_{1,2}& b_{1,3}& \ldots & b_{1,n}\\
  \vdots & \vdots & \vdots &\ddots & \vdots \\
  a_{n} & b_{n,2}& b_{n,3}& \ldots & b_{n,n}
 \end{array}\right|.
 \end{split}
\end{equation*}

	\section{Problem formulation}
	
 	Consider the generic signal detection problem in colored Gaussian noise: $\mathbf{x}=\sqrt{\rho}\mathbf{h},  s+\mathbf{n}$
	where $\mathbf{x,h}\in\mathbb{C}^{m}$, $\rho>0$, $s\sim\mathcal{CN}(0,1)$ and  $\mathbf{n}\sim \mathcal{CN}_m(\mathbf{0}, \boldsymbol{\Sigma})$. Here the noise covariance matrix $\boldsymbol{\Sigma}$ may be known or unknown at the detector. The classical signal detection problem can be formulated as the following hypothesis testing problem
	\begin{align*}
	&\mathcal{H}_0:\; \rho=0\;\;\;\;\;\; \text{Signal is absent}\\
	& \mathcal{H}_1:\; \rho>0 \;\;\;\;\; \text{Signal is present}.
	\end{align*}
	Nothing that the covariance matrix of $\mathbf{x}$ can be written as $\mathbf{S}=\rho \mathbf{h}\mathbf{h}^\dagger+\boldsymbol{\Sigma}$,
	where $(\cdot)^\dagger$ denotes the conjugate transpose, we can have the following equivalent form
	\begin{align*}
	\begin{array}{ll}
	\mathcal{H}_0:\; \mathbf{R}=\boldsymbol{\Sigma} &\text{Signal is absent}\\
	\mathcal{H}_1:\; \mathbf{S}=\rho\mathbf{h}\mathbf{h}^\dagger+\boldsymbol{\Sigma} & \text{Signal is present}.
	\end{array}
	\end{align*}
	Let us now consider the matrix $\boldsymbol{\Psi}=\mathbf{R}^{-1}\mathbf{S}$ with the eigenvalues $\lambda_1\leq \lambda_2\leq \ldots\leq \lambda_m$. As such we have
	\begin{align*}
	\boldsymbol{\Psi}=\mathbf{R}^{-1}\mathbf{S}=
	\rho \boldsymbol{\Sigma}^{-1}\mathbf{h}\mathbf{h}^\dagger+\mathbf{I},
	\end{align*}
	from which we can observe that in the presence of a signal, the maximum eigenvalue of $\boldsymbol{\Psi}$ (i.e., $\lambda_m$) is strictly greater than one, whereas the other eigenvalues are equal to one (i.e., $\lambda_1=\lambda_2=\ldots=\lambda_{m-1}=1$). Capitalizing on this observation Rao and Silverstein \cite{Nadakuditi2010jsacsp} concluded that, given the knowledge of $\mathbf{R}$ and $\mathbf{S}$, the maximum eigenvalue of $\boldsymbol{\Psi}$ could be used to detect the presence of a signal. It is noteworthy that the matrix $\boldsymbol{\Psi}$ is also known as the $F$-matrix or Fisher matrix. 
	
	In most practical settings,  $\mathbf{R}$ and $\bf{S}$ matrices  are unknown. To circumvent this difficulty, it is common to replace $\mathbf{R}$ and $\mathbf{S}$ by their sample estimates. To this end, let us assume that we have $p > 1 $ i.i.d. sample observations from signal-plus-noise scenario given by $\{\mathbf{x}_1, \mathbf{x}_2,\ldots, \mathbf{x}_p\}$, and $n$ i.i.d. sample observations from noise-only scenario given by $\{\mathbf{n}_1, \mathbf{n}_2,\ldots,\mathbf{n}_n\}$. Thus,  the  sample estimates of $\mathbf{R}$ and $\mathbf{S}$ become 
	\begin{align}
	\widehat{\mathbf{R}}=\frac{1}{n}\sum_{\ell=1}^n \mathbf{n}_\ell \mathbf{n}_\ell^\dagger
	\quad \text{and}\quad
	\widehat{\mathbf{S}}=\frac{1}{p}\sum_{k=1}^p \mathbf{x}_k\mathbf{x}_k^\dagger
	\end{align} 
	where we assume that $n,p\geq m$ (this ensures that both $\widehat{\mathbf{R}}$ and $\widehat{\mathbf{S}}$ are positive definite with probability $1$ \cite{memuirhead2009aspects}). Consequently, following Rao \cite{Nadakuditi2010jsacsp}, we form the matrix
	\begin{align}
	\widehat{\boldsymbol{\Psi}}=\widehat{\mathbf{R}}^{-1}\widehat{\mathbf{S}}
	\end{align}
	and focus on its maximum eigenvalue as the test statistic.\footnote{This is also known as the Roy's largest root test which is a consequence of Roy's union intersection principle \cite{Mardia1979book}.}. As such, we have $
	\widehat{\mathbf{R}}\sim\mathcal{CW}_m\left(n, \boldsymbol{\Sigma}\right)$ and $
	p\widehat{\mathbf{S}}\sim\mathcal{CW}_m\left(p, \boldsymbol{\Sigma}+\rho \mathbf{h}\mathbf{h}^\dagger\right)
	$.
	Noting that the eigenvalues of $ \widehat{\boldsymbol{\Psi}}$ do not change under the simultaneous transformations $\widehat{\mathbf{R}}\mapsto \boldsymbol{\Sigma}^{-1/2}\widehat{\mathbf{R}}\boldsymbol{\Sigma}^{-1/2}$, and $\widehat{\mathbf{S}}\mapsto \boldsymbol{\Sigma}^{-1/2}\widehat{\mathbf{S}}\boldsymbol{\Sigma}^{-1/2}$, without loss of generality we assume that
	$\boldsymbol{\Sigma}=\sigma^2\mathbf{I}_m$. Therefore, in what follows we focus on the maximum eigenvalue of  $ \widehat{\boldsymbol{\Psi}}$, where
	\begin{align}
	&n\widehat{\mathbf{R}}\sim\mathcal{CW}_m\left(n, \mathbf{I}_m\right)\\
	& p\widehat{\mathbf{S}}\sim\mathcal{CW}_m\left(p, \mathbf{I}_m+\gamma \mathbf{u}\mathbf{u}^\dagger\right)
	\end{align}
	with $\gamma=\rho ||\mathbf{h}||^2/\sigma^2$ and $\mathbf{u}=\mathbf{h}/||\mathbf{h}||$ being a unit vector.   
	
	Let us denote the maximum eigenvalue of $ \widehat{\boldsymbol{\Psi}}$ as $\hat{\lambda}_{\max}(\gamma)$. Now, in order to assess the performance of the maximum-eigen based  detector, we need to evaluate the detection\footnote{This is also known as the power of the test.} and false alarm probabilities. They may be expressed as
		\begin{align}
	& P_D(\gamma, \mu)=\Pr\left(\hat{\lambda}_{\max}(\gamma)>\mu_{\text{th}}|\mathcal{H}_1\right) \label{detection}\\
	& P_F(\gamma,\mu)=\Pr\left(\hat{\lambda}_{\max}(\gamma)>\mu_{\text{th}}|\mathcal{H}_0\right) \label{alarm}
	\end{align}
	where $\mu_{\text{th}}$ is the threshold. The $(P_D, P_F) $ characterizes   the detector  and is called the  ROC profile.
	
	The main challenge here is to characterize the maximum eigenvalue of $\widehat{\boldsymbol{\Psi}}$ under the alternative $\mathcal{H}_1$. To this end, in this paper, we use orthogonal polynomial techniques due to Mehta \cite{me11} to obtain a closed form solution to this problem. In particular, we derive an expression which contains a determinant whose dimension depends through the relative difference between $m$ and $n$

\section{C.D.F. of the Maximum Eigenvalue}
    
    Before proceeding further, we present some fundamental results pertaining to the  joint eigenvalue distribution of an $F$-matrix and Jacobi polynomials.
    
    \subsection{Preliminaries}
\begin{defn}\label{F_Def}
Let $\mathbf{W}_1\sim\mathcal{W}_m\left(p,\boldsymbol{\Sigma}\right)$ and $\mathbf{W}_2\sim\mathcal{W}_m\left(n,\mathbf{I}_m\right)$ be two independent Wishart matrices with $p,n\geq m$. Then the joint eigenvalue density of the ordered eigenvalues, $\lambda_1\leq \lambda_2\leq \ldots\leq \lambda_m$, of $\mathbf{W}_1\mathbf{W}_2^{-1}$ is given by \cite{meJames}
\begin{align}
\label{james_joint}
f(\lambda_1,\cdots,\lambda_m)& =\frac{\mathcal{K}_1(m,n,p) }{\text{det}^p\left(\boldsymbol{\Sigma}\right)}
	\prod_{j=1}^m \lambda_j^{p-m}\Delta_m^2(\boldsymbol{\lambda})\nonumber\\
	& \qquad\times {}_1\widetilde F_0\left(p+n;-\boldsymbol{\Sigma}^{-1}, \boldsymbol{\Lambda}\right)
\end{align}
where  ${}_1\widetilde F_0\left(\cdot;\cdot,\cdot\right)$ is the generalized complex hypergeometric function of two matrix arguments, $\Delta_m^2(\boldsymbol{\lambda})=\prod_{1\leq i<j\leq m}\left(\lambda_j-\lambda_i\right)$ is the Vandermonde determinant, $\boldsymbol{\Lambda}=\text{diag}\left(\lambda_m,\ldots,\lambda_1\right)$, and 
$
\mathcal{K}_1(m,n,p)=\frac{\pi^{m(m-1)}\widetilde{\Gamma}_m(n+p)}{\widetilde{\Gamma}_m(m)\widetilde{\Gamma}_m(n)\widetilde{\Gamma}_m(p)}
$
with the complex multivariate gamma function is written in terms of the classical gamma function $\Gamma(\cdot)$ as
$
\widetilde{\Gamma}_m(n)=\pi^{\frac{1}{2}m(m-1)} \prod_{j=1}^m \Gamma\left(n-j+1\right).
$

\end{defn}

\begin{defn}
Jacobi polynomials can be defined as follows \cite[eq. 5.112]{me12}
	\begin{equation}\label{jacobidef1}
	P_{n}^{(a,b)}(x) = \sum_{k=0}^{n}\binom{n+a}{n-k}\binom{n+k+a+b}{k}\left(\frac{x-1}{2}\right)^{k} \hspace{6mm} 
	\end{equation} 
    
    where $a,b>-1$, $\binom{n}{k}=\frac{n!}{(n-k)! k!}$ with $n\geq k\geq 0$.
    \end{defn}
 \subsection{Finite Dimensional Analysis of the C.D.F.}
Having defined the above preliminary quantities, now we focus on deriving a new c.d.f. for the maximum eigenvalue of $\mathbf{W}_1\mathbf{W}_2^{-1}$ when the covariance matrix $\boldsymbol{\Sigma}$ takes the so called rank-$1$ spiked form. In this case, the covariance matrix can be decomposed as
\begin{align}
\label{spike_cov}
\boldsymbol{\Sigma}=\mathbf{I}_m+\eta \mathbf{vv}^\dagger=\mathbf{V}\text{diag}\left(1+\eta, 1, 1,\ldots, 1\right)\mathbf{V}^\dagger 
\end{align}
where $\mathbf{V}=\left(\mathbf{v}\; \mathbf{v}_2\; \ldots \mathbf{v}_m\right)\in\mathbb{C}^{m\times m}$ is a unitary matrix and  $\eta\geq 0$. 
Following Khatri \cite{Khatri}, the hypergeometric function of two matrix arguments given in the join density (\ref{james_joint}) can be written as a ratio between the determinants of two $m\times m$ square matrices. Since the eigenvalues of the matrix $\boldsymbol{\Sigma}^{-1}$ are such that $1/(1+\eta)$ has algebraic multiplicity one  and $1$ has algebraic multiplicity $m-1$, the resultant  ratio takes an indeterminate form. Therefore, one has to repeatedly apply L\^ospital's rule to obtain a deterministic expression. However, that expression is not amenable to apply Mehta's \cite{me11} orthogonal polynomial technique. Therefore, in view of applying the powerful orthogonal polynomial technique, we derive an alternative expression for the joint eigenvalue density. This alternative derivation techniques has also been used earlier in \cite{Dharmawansa2014stat} to derive a single contour integral representation for the joint eigenvalue density when the matrices are real\footnote{It is noteworthy that when the matrices are real, the hypergeometric function of two matrix arguments does not admit such a determinant representation.}. The following corollary gives the alternative expression for the joint density.

 \begin{cor}\label{joint_eig_pdf}
 Let $\mathbf{W}_1\sim\mathcal{W}_m(p,\mathbf{I}_m+\eta \mathbf{v}\mathbf{v}^\dagger)$ and $\mathbf{W}_2\sim\mathcal{W}_m(n,\mathbf{I}_m)$ be independent Wishart matrices with $m\leq p,n$ and $\eta\geq 0$. Then the joint density of the ordered eigenvalues $0\leq \lambda_1\leq \lambda_2\leq\cdots\leq \lambda_m<\infty$ of $\mathbf{W}_1\mathbf{W}_2^{-1}$ is given by
		\begin{align}
		\label{jpdf}
		f(\lambda_1,\cdots,\lambda_m)=
		f_{\text{uc}}(\lambda_1,\cdots,\lambda_m) f_{\text{cor}}(\lambda_1,\cdots,\lambda_m)
		\end{align}
		where
		\begin{align*}
		f_{\text{uc}}(\lambda_1,\cdots,\lambda_m)=\mathcal{K}_1(m,n,p)
		\prod_{j=1}^m \frac{\lambda_j^{p-m}}{(1+\lambda_j)^{p+n}} \Delta_m^2(\boldsymbol{\lambda}),
		\end{align*}
		\begin{align*}
		f_{\text{cor}}(\lambda_1,\cdots,\lambda_m)&=\frac{\mathcal{K}_2(m,n,p)}{\eta^{m-1}(1+\eta)^{p+1-m}}
		\prod_{j=1}^m (1+\lambda_j)\\
		& \times 
		\sum_{k=1}^m
		\frac{(1+\lambda_k)^{p+n-1}}{\displaystyle\prod_{\substack{j=1\\
					j\neq k}}^m(\lambda_k-\lambda_j)
		\left(1+\frac{\lambda_k}{\eta+1}\right)^{p+n+1-m}},
		\end{align*}
        and 
		$\mathcal{K}_2(m,n,p)=\frac{(m-1)!(p+n-m)!}{(p+n-1)!}$.
 \end{cor}
\begin{IEEEproof} 
Omitted due to space limitations.
\end{IEEEproof}
        \begin{rk}
		It is worth noting that the function $f_{\text{uc}}(\lambda_1,\lambda_2,\cdots,\lambda_m)$ denotes the joint density of the ordered eigenvalues of $\mathbf{W}_1\mathbf{W}_2^{-1}$ corresponding to the case $\mathbf{W}_1\sim\mathcal{W}_m(p,\mathbf{I}_m)$ and $\mathbf{W}_2\sim\mathcal{W}_m(n,\mathbf{I}_m)$.
	\end{rk}
	\begin{rk}
		Alternatively, the above expression can be used to obtain the joint density of the ordered eigenvalues of {\it deformed Jacobi ensemble}, $\mathbf{W}_1(\mathbf{W}_2+\mathbf{W}_1)^{-1}$ with $\mathbf{W}_1\sim\mathcal{W}_m(p,\mathbf{I}_m+\eta \mathbf{vv}^\dagger)$ and $\mathbf{W}_2\sim\mathcal{W}_m(n,\mathbf{I}_m)$.
	\end{rk}
	
   We may use the above join density to obtain the  c.d.f. of the maximum eigenvalue, which is given by the following theorem.

	\begin{thm}\label{alter}
		Let $\mathbf{W}_1\sim\mathcal{W}_m(p,\mathbf{I}_m+\eta \mathbf{vv}^\dagger)$ and $\mathbf{W}_2\sim\mathcal{W}_m(n,\mathbf{I}_m)$ be independent with $m\leq p,n$ and $\eta\geq 0$. Then the c.d.f. of the maximum eigenvalue $ \lambda_{\max} $ of $\mathbf{W}_1\mathbf{W}_2^{-1}$ is given by
		\begin{align}\label{cdfthm}
			F^{(\alpha)}_{\lambda_{\max}}(t;\eta)
			&=\dfrac{\mathcal{K}(m,p,\alpha)}{(p-1)!(1+\eta)^{p}}\left(\dfrac{t}{1+t}\right)^{m(\alpha+\beta+m)}\nonumber\\
			& \quad \times 
			\det\left[\Phi_{i}(t,\eta)\hspace{3mm} \Psi_{i,j}(t)\right]_{\substack{i=1,2,...,\alpha+1\\j=2,3,...,\alpha+1}}
		\end{align}
		where
        \begin{dmath*}
			\Psi_{i,j}(t)= (m+i+\beta-1)_{j-2}P_{m+i-j}^{(j-2,\beta+j-2)}\left(\frac{2}{t}+1\right),
		\end{dmath*}
		\begin{align*}
		&\Phi_{i}(t,\eta)\\
		& =\mathcal{Q}_i(m,n,p)\sum_{k=0}^{\alpha-i+1}\frac{(p+i-1)_k(\alpha-i+2)!}{k!(p+m+2i-2)_k(\alpha-i-k+1)!}\\
		& \qquad \qquad \qquad \qquad \qquad \times \frac{\left(\eta t\right)^{k+i-1}\left((1+\eta)(1+t)\right)^{p+k}}{\left(1+\eta +t\right)^{p+k+i-1}},
		\end{align*}
		$
			\mathcal{Q}_i(m,n,p)=\frac{(n+p+i-2)!(p+i-2)!}{(p+m+2i-3)!},
	$
        and 
	$
        \mathcal{K}(m,p,\alpha)=\prod_{j=0}^{\alpha-1}\frac{(p+m+j-1)!}{(p+m+2j)!}
       $
		with $\alpha = n-m$ and $\beta=p-m$.	
	\end{thm}
	\begin{IEEEproof}
	Omitted due to space limitations.
	\end{IEEEproof}
	
  The new exact c.d.f. expression for the maximum eigenvalue of $\mathbf{W}_1\mathbf{W}_2^{-1}$, which contains the determinant of a square matrix whose dimension depends on the difference $\alpha=n-m$, is highly desirable when the difference between $m$ and $n$ is small irrespective  of their individual magnitudes. For instance, when $n=m$ (i.e., $\alpha=0$) the determinant vanishes and we obtain a scalar result as shown below. This is one of the many advantages of using the orthogonal polynomial approach. This key representation, also facilitates the derivation of the limiting eigenvalue distribution of the maximum eigenvalue (i.e., the limit when $m,n\to\infty$ such that $m-n$ is fixed).
  
\begin{cor}
		The exact c.d.f. of the maximum eigenvalue of $\mathbf{W}_1\mathbf{W}_2^{-1}$corresponding to $\alpha = 0$ is given by
		\begin{dmath}\label{cdfalpha0}
			F^{(0)}_{\lambda_{\max}}(t;\eta)
			=\left(\dfrac{t}{1+t}\right)^{mp}\left(1+\dfrac{\eta }{1+t}\right)^{-p}.
		\end{dmath}
	\end{cor}  
    
    Having armed with the above characteristics of the maximum eigenvalue of $\mathbf{W}_1\mathbf{W}_2^{-1}$, in the following section, we focus on the ROC of the maximum eigenvalue based detector.

    \section{ROC of the Maximum Eigenvalue of $\widehat{\bf{\Psi}}$}
    Let us now investigate the behavior of detection and false alarm probabilities associated with the maximum eigenvalue based test. To this end, noting that the eigenvalues of $\widehat{\boldsymbol{\Psi}}$ and $\mathbf{W}_1\mathbf{W}_2^{-1}$ are related by $\hat{\lambda}_j=(n/p)\lambda_j$, for $j=1,2,\ldots,m$, we can represent the c.d.f. of the maximum eigenvalue corresponding to $\widehat{\boldsymbol{\Psi}}$ as
    $F_{\lambda_\max}^{(\alpha)}(\kappa x;\gamma)$, where $\kappa=p/n$.

    Now following Theorem \ref{alter} along with with (\ref{detection}), (\ref{alarm}), the detection and false alarm probabilities can be written, respectively, as
    \begin{align}
    P_D(\gamma, \mu_{\text{th}})&=1-F_{\lambda_\max}^{(\alpha)}(\kappa\mu_{\text{th}};\gamma)\\
    P_F(\mu_{\text{th}})&=1-F_{\lambda_\max}^{(\alpha)}(\kappa\mu_{\text{th}};0). 
    \end{align}
    In general, deriving a functional relationship between $P_D$ and $P_F$ by eliminating the parametric dependency on $\mu_{\text{th}}$ is an arduous task. However, when $\alpha$ admits zero, we can obtain an explicit relationship between them as shown in the following corollary.
    \begin{cor}
    \label{corasybalanced}
    Let us suppress the parameters $\gamma$, $\mu_{\text{th}}$ and represent the detection and false alarm probabilities, respectively as $P_D$ and $P_F$. Then, when $\alpha=0$, we have the following functional relationship between $P_D$ and $P_F$
    \begin{align}
    \label{rocbalanced}
    P_D=1-\frac{1-P_F}{\left(1+\gamma-\gamma\left[1-P_F\right]^{1/mp}\right)^p}.
    \end{align}
    \end{cor}

From the above relation, taken $P_D$ as a function of $\gamma$, we can easily see that, for $\gamma_1>\gamma_2$, $P_D(\gamma_2)>P_D(\gamma_1)$. 
This conforms the common observation that the SNR is positively correlated with the detection probability for a fixed value of $P_F$.
\begin{figure}[ht]
		\centering
		\includegraphics[width=0.4\textwidth]{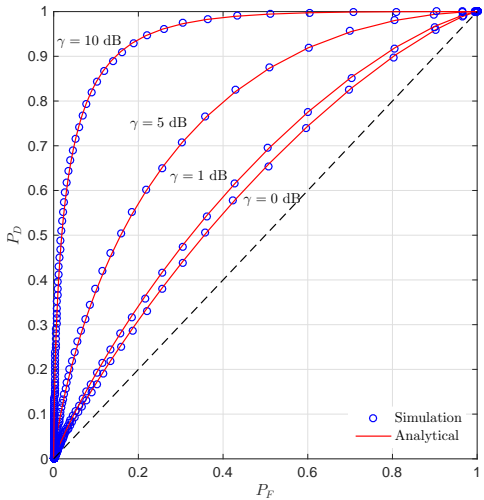}
		\caption{$P_D$ vs $P_F$ for different valued of $\gamma$ with $(m,n,p)=(5,8,10)$.}
		\label{ROCscene2a}
	\end{figure} 
\begin{figure}[ht]
		\centering
		\includegraphics[width=0.4\textwidth]{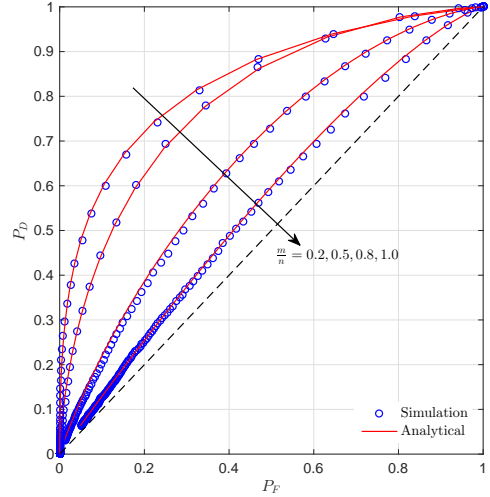}
		\caption{$P_D$ vs $P_F$ for different valued of  $m/n$ with $m/p=1$ and $n=10$ when $\gamma=5$\,dB.}
		\label{ROCscene3a}
	\end{figure} 

The ROC curves corresponding to different parameter settings are shown in Figs. \ref{ROCscene2a} and \ref{ROCscene3a}. 
The ROC of the maximum eigenvalues is shown in Fig. \ref{ROCscene2a} for different SNR values. The ROC improvement with the increasing SNR is clearly visible in Fig. \ref{ROCscene2a}. The next important frontier which affects the ROC profile is the dimensionality of the matrices. Therefore, let us now numerically investigate the effect of the matrix dimensions on the ROC profile. To this end, Fig. \ref{ROCscene3a} shows the effect of $m/n$ for $m/p=1$. As can be seen, the disparity between $m$ and $n$ improves the ROC profile. The reason behind this observation is that the quality of the sample covariance matrix is improved when the length of the data record (i.e.,$n$) increases in comparison with the dimensionality of the receiver (i.e., $m$). Since the minimum requirement for  $\bf{\widehat{R}}$ to be invertible is $m=n$, we can observe the worst ROC performance corresponds to $m/n=1$. 

The joint effect of $m$ and $p$ is characterized with respect to the scenario where  $m$ and $p$
 both vary such that $m/p=\nu>0$ is constant. After some algebra, we conclude that $P_D$ attains its maximum at $p=p^*$ ($m^*=\nu p^*$), where
\begin{align}
\sqrt{\frac{-\ln(1-P_F)}{-2\nu\ln\left(\frac{\gamma+1}{\gamma+2}\right)}}<p^*<\sqrt{\frac{-\ln(1-P_F)}{-\nu\ln\left(\frac{\gamma+2}{\gamma+4}\right)}}.
\end{align}
Having obtained the upper and lower bounds on $p^*$, a good approximation of $p^*$ can be written as\footnote{In general any convex combination of the upper and lower bounds can be a candidate for the  $p^*$.}
\begin{align}
\label{eq pmax approx}
p^*\approx \frac{1}{2}\left(\sqrt{\frac{-\ln(1-P_F)}{-\nu\ln\left(\frac{\gamma+2}{\gamma+4}\right)}}+\sqrt{\frac{-\ln(1-P_F)}{-2\nu\ln\left(\frac{\gamma+1}{\gamma+2}\right)}}\right).
\end{align}
\begin{figure}[t]
		\centering
		\includegraphics[width=0.4\textwidth]{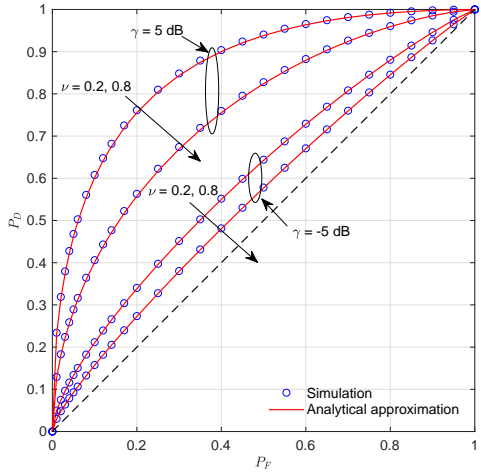}
		\caption{$P_D$ vs $P_F$ for the optimal $p$ and approximated $p$.}
		\label{p_approxa}
	\end{figure}
The above process suggests us that when $m$ and $p$ diverge such that their ratio approaches a certain limit, the maximum eigenvalue gradually loses its power.

To further highlight the accuracy of the proposed approximation, in Fig.~\ref{p_approxa} we compare the optimal ROC profiles evaluated based on (\ref{eq pmax approx}) and by numerically optimizing (\ref{rocbalanced}). As can be seen from the figure, the disparity between the proposed approximation and the exact optimal solution is insignificant. Therefore, when $m=n$, under the second scenario, we can choose $p$ as per (\ref{eq pmax approx}) for fixed $P_F$, $\gamma$, and $\nu$ in view of maximizing the detection probability.        




    



\vspace{-1mm}
\section{Conclusion}
This paper investigates the signal detection problem in colored noise with unknown covariance matrix. In particular, we focus on detecting the presence of a signal using the maximum generalized eigenvalue of so called whitened sample covariance matrix. Therefore, the performance of this detector amounts to determining the statistics of the maximum eigenvalue of the deformed JUE. To this end, following the powerful orthogonal polynomial approach, we have developed a new expression for the c.d.f. of the maximum  eigenvalue of the deformed JUE. 
We then use this new c.d.f. expression to determine the ROC of the detector. It turns out that, for a fixed SNR, when $m$ (i.e., the dimensionality of the detector), $n$ (i.e., the number of noise-only samples), and $p$ (i.e., the number of signal-plus-noise samples) increase over finite values such that $m=n$ and $m/p$ is constant, we obtain an optimal ROC profile corresponding to specific $m,n$, and $p$ values. Therefore, in the above setting, when $m,p$, and $n$ increase asymptotically, the maximum eigenvalue gradually loses its detection power. This is not surprising, since under the above asymptotic setting, the detector operates below the so called phase transition where the maximum eigenvalue has no detection power.

   
	\appendices
	
	\ifCLASSOPTIONcaptionsoff
	\newpage
	\fi


	

	\end{document}